\documentclass[preprint,showpacs,preprintnumbers,amsmath,amssymb]{revtex4}

\usepackage{graphicx}
\usepackage{dcolumn}
\usepackage{bm}

\begin{document}

\title{Modeling of quasiparticle-induced excitations of a Josephson charge-phase qubit}

\author{J. K\"{o}nemann, H. Zangerle, B. Egeling,  R. Harke, B. Mackrodt, R. Dolata, and A.~B. Zorin}

 \affiliation{Physikalisch-Technische Bundesanstalt, Bundesallee 100, 38116 Braunschweig, Germany}%

\date{\today}

\begin{abstract}
We have analyzed quasiparticle transitions in an Al charge-phase
qubit inducing a dynamic change of the qubit states. The
time-averaged mixed state is related to the strong coupling of the
qubit to an ensemble of non-equilibrium quasiparticles in the
leads. Such quasiparticles tunnel stochastically on and off the
island and can excite the qubit. Continuous monitoring of the
qubit impedance at a frequency of $80$ MHz shows the admixture of
the excited state. We present a numerical description
of these cyclic transitions and compare it with our experimental
data.
\end{abstract}

\maketitle

Superconducting circuits, based on small Josephson junctions and
tunneling of single Cooper pairs, are promising candidates for
qubits \cite{makhlin}.  The qubit operation relies on the coherent
superposition of the macroscopic quantum states of the qubit
circuits, but the incoherent tunneling of a single unpaired
electron, i.e. a quasiparticle (QP), changes
instantly and stochastically the even-parity to an odd-parity
charge state  in the system \cite{qubit-Saclay,qubit-Chalmers} and thus shifts
 the bias point of the qubit. Moreover, it halts the coherent
tunneling of Cooper pairs in a random telegraphic way,  being
a fundamental source of decoherence \cite{ieee,glazman2} and hence
setting limits to the qubit operation time.

Several experimental techniques have been applied to observe the
even-odd states of superconducting circuits, basically either
focusing on dc measurements of the superconducting branch of
single-charge transistors \cite{tuominen,Amar,chargeexp},
electrometry of the charge of Cooper pair boxes
\cite{Bouchiat,lehnert,lukens,echternach}, or by rf techniques with
inductively-coupled resonant circuits, i.e. by the
rf reflectrometry method  imaging the effective
Josephson-inductance of the device \cite{aumentado}. This
rf reflectrometry method has been applied recently to investigate
the kinetics of quasiparticle tunneling
\cite{clark,echt,wellstood}
 and QP trapping in Cooper pair boxes \cite{lutexp}.

In this paper we focus on the problem of single QPs entering the
qubit island. By continuous monitoring of the qubit impedance at
a frequency of 80 MHz, we exploit the quasiparticle-induced dynamic
change of the qubit states. These quasiparticle transitions induce
a time-averaged mixed qubit state related to the strong coupling
of the qubit to an ensemble of non-equilibrium QPs in the leads.
In a previous paper we  reported this effect and showed, that these transitions
obey a selection rule \cite{kong}. Here we focus on the kinetics
of the cyclic quasiparticle transitions in our circuit and present
a numerical modeling of these transitions.

In our  experiment,  we have investigated a Josephson charge-phase
qubit \cite{zang,Zorin-JETP,cpqb2,imt2}. This device can be
considered as a Cooper-pair box of SQUID configuration, i.e. a
superconducting loop interrupted by two small-capacitance
Josephson-junctions with a mesoscopic island in between. This island is
capacitively coupled to a gate electrode, see Fig.\,1 (a). The
 quantum states $|n,q\rangle$  of  our qubit system are associated with different Bloch bands
  of a
particle in the periodic (Josephson) potential \cite{sqc}. Here
$n$ is the band number and $q$ the quasicharge governed by the
gate voltage $V_{\rm G}$, i.~e.~$q=C_{\rm G}V_{\rm G}$, where
$C_{\rm G}$ is the gate capacitance, see, e.g.,
Ref.\,\cite{CP-R-electr-theo}. The quantum states of the
transistor also
 involve the phase coordinate $\varphi_{\rm dc}$ set by the external
magnetic flux $\Phi_{\rm dc}$ applied to the SQUID loop.  The
qubit
  Josephson inductance $L_{\rm J}$ (being much larger than the inductance $L$ of the loop) is related to the
local curvature of the corresponding energy surface $E_{\rm
n}(q,\varphi)$, i.e. $1/L_{\rm J}(n,q,\varphi)\propto
\partial^2E_{\rm n}(q,\varphi)/\partial\varphi^2$, where the integer values $n=0$ and $1$ correspond to the ground and excited state,
respectively.

Our Josephson qubit is inductively coupled to an rf-driven tank
circuit. The qubit eigenstates can be identified by means of the
Josephson inductance of the transistor which is probed by small
rf oscillations $I_{\rm RF}$ in the loop with the resulting phase
$\varphi(t)=\varphi_{\rm dc}+a\sin{(2\pi ft)}$. Here, $a$ is
proportional to the amplitude of the rf oscillations in the tank
circuit induced by an rf driving current of frequency $f$, close to
the bare resonance frequency $f_0 \approx 77$\,MHz of the tank
circuit.  Due to the coupling to the qubit, the effective
inductance $L_{\rm eff}$ of the  circuit is equal to
$L_{\rm T}-M^2L_{\rm J}^{-1}(n,q,\varphi)$, with the geometrical
inductance $L_{\rm T}\approx 150$ nH of the resonant circuit, with
the mutual inductance $M\approx 3.8$ nH, and with  $k\approx 0.4$
being the coupling coefficient between the superconducting qubit
loop and the resonant circuit (for details, see Ref. \cite{kong}).
The resulting shift $\Delta f$ of the resonant frequency is
$\Delta f/f_0\propto -1/L_{\rm J}(q,\varphi).$

The investigated sample has been fabricated in two steps: First,
the tank circuit inductor was fabricated on the basis of Nb
technology, including e-beam lithography, dry etching, anodization
and chemical-mechanical polishing, see Refs.
\cite{dolata05,zang} for details. In a second step, the qubit loop and the
Bloch transistor were co-fabricated on the same chip by means of
the two-angle Al shadow evaporation technique. In such an rf-SQUID
configuration, the Bloch transistor is galvanically decoupled from
the measuring circuit, which in general leads to a reduction in
the density of non-equilibrium QPs which are able to enter the
island. Besides this, no further precautions for the suppression
of QP poisoning of the island - such as, for example, the
engineering of a barrier-like gap profile having the island gap
value  greater than the electrode gap value
\cite{chargeexp3,tsai}, or the implementation of normal-metal QP
traps \cite{chargeexp} in the outer electrodes - were taken.  The
 critical currents of the individual junctions of the transistor were approx.
25 nA, with the corresponding value of 45  $\mu$eV for the average
Josephson coupling energy $E_{\rm J0}$. The charging energy
$E_{\rm C}$ of the transistor island has a value of 110 $\mu$eV,
such that both energies $E_{\rm C}$ and $E_{\rm J0}$ are smaller
than the value of the superconducting gap in Al films,
$\Delta_{\rm Al}\approx 210$ $\mu$eV.  These values of $E_{\rm
J0}$ and $E_{\rm C}$ were taken from a fitting of the shape of the
ground state extracted from rf measurements with a finite amplitude
of the Josephson phase oscillations, see Ref.\cite{kong} for
details.

\begin{figure}
\includegraphics[width=0.89\linewidth]{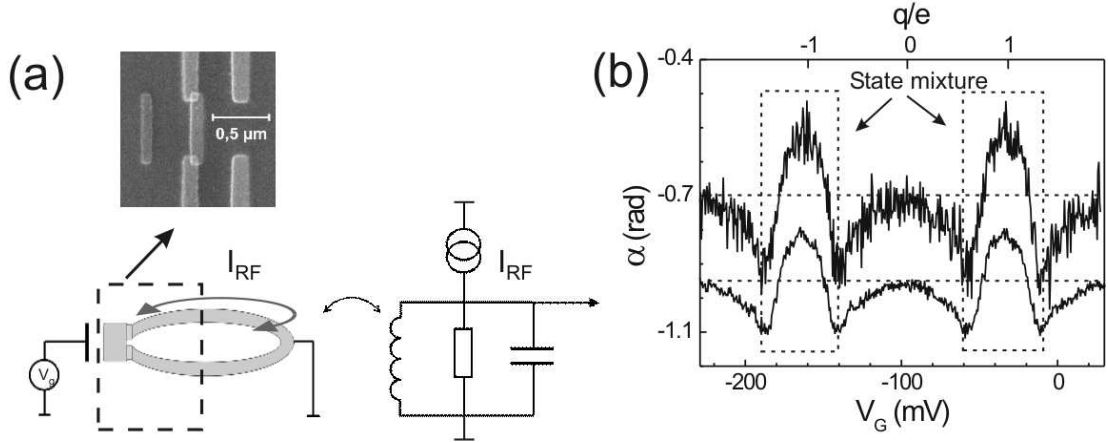}
\caption{\label{fig:fig1} (a) Sketch of the charge-phase qubit
with inductive readout.
 The core element is a double Josephson junction with a capacitive gate coupled to its island, i.e. the
Bloch transistor, embedded in a macroscopic superconducting loop.
The loop is inductively coupled to an rf-driven tank circuit which
itself is capacitively coupled to a cold preamplifier, see Ref.
\cite{kong}. The inset shows a scanning electron micrograph of the
transistor island, fabricated in the Al shadow evaporation technique.
(b) Experimental gate modulation dependencies $\alpha(q)$ for an
amplitude of rf oscillations $a=0.28$ (top curve) and $a=0.84$
(bottom curve). The range of gate voltage shown by dashed-line boxes corresponds to the mixing of ground and excited states.}
\end{figure}
In our experiment, which was carried out in a dilution
refrigerator at a base temperature of 20 mK, we measure  the phase
angle $\alpha$ between the driving signal $I_{\rm rf}$ and the rf
voltage $V_{\rm rf}$ on the tank circuit instead of the frequency
detuning $\Delta f$. From the $\alpha$-dependence one can deduce
the Josephson-inductance $L_{\rm J}(q,\varphi)$ by applying the
simple formula \cite{zang}:
\begin{equation}
\tan{\alpha}=k^2Q\frac{L}{L_{\rm J}(n, q, \varphi)}. \label{fa}
\end{equation}

The measurement of the phase shift, $\alpha$ as a function of
$\varphi_{\rm dc}$ and  $q$ allows the curvature of the energy
surface to be mapped, showing a periodical dependence of $\alpha$
both on $\varphi_{\rm dc}$ with a period of $2\pi$ and $q\propto
 V_{\rm G}$ with a
$2e$-periodic gate modulation. In this paper, we focus on the measurements carried out at $\varphi_{\rm dc}=\pi$ that correspond to the minimum qubit energy (in the degeneracy point $q=e$ equal to the difference of the Josephson coupling energies of the individual junctions).
In this case, the effect of the quasiparticle induced transitions, discussed in the following, is strongest. These transitions manifest themselves 
in an overshooting behavior of the gate dependence curve
 measured at $\varphi_{\rm dc}=\pi$. This peculiar shape is formed by two types of arcs, see Fig.\,1 (b). The
obtuse arcs, for example,  (between the degeneracy points at $q=-e$ and $q=e$) are
interrupted by the acute arcs centered at the degeneracy points, i.e. the
phase angle $\alpha$ starts to rise sharply  and remains in a
broad range around $q=\pm e$ at a level that is even higher than
that for $q=0$ (indicated by the dotted lines in Fig.\,1 (b)).
When increasing the amplitude $a$ of the rf oscillations  of the
Josephson phase, the Josephson-inductance, i.e. the second derivative of the energy surface, is
probed over a larger region of $\varphi$. Thus, the gate curve is averaged over states with different $E_{\rm
J}(\varphi)=\sqrt{E_{J1}^2+E_{J2}^2+2E_{J1}E_{J2}\cos{\varphi}}$ for fixed $q$ with $E_{J1,2}$ being the Josephson energies of the individual tunneling junctions. As a consequence, the overshooting is reduced due to the increased effective splitting between  ground-state and excited state.

 We  explain this overshooting behavior by a
statistical mixture of the different quantum states of the qubit
 ground and excited state that are characterized by an opposite curvature of the corresponding energy bands \cite{Zorin-JETP}. Here, at the degeneracy point ($q=e$ and
$\varphi_{\rm dc}=\pi$), the states considered are the ground
state $A$ and the excited state $B$, both with even parity of the
island. We can neglect the contribution of the ground state $C$
with odd parity, as the admixture of the values of $\alpha(q)$
corresponding to this state with small negative curvature cannot
yield the above-mentioned overshooting of the experimental data in
the vicinity of $q=e$. Likewise, we are of the opinion that the
contribution of the odd state is small due to the presumably short
lifetime of a QP in the island, see, e.g.,
Refs.\cite{glazman2,lutexp}. Besides, we do not consider the
corresponding odd-parity excited state (not marked here), since
its excitation energy ($\approx 3E_{\rm C}$) is too great, i.e. much
larger than the energy gap between the states $A$ and $B$.

\begin{figure}
\includegraphics[width=0.89\linewidth]{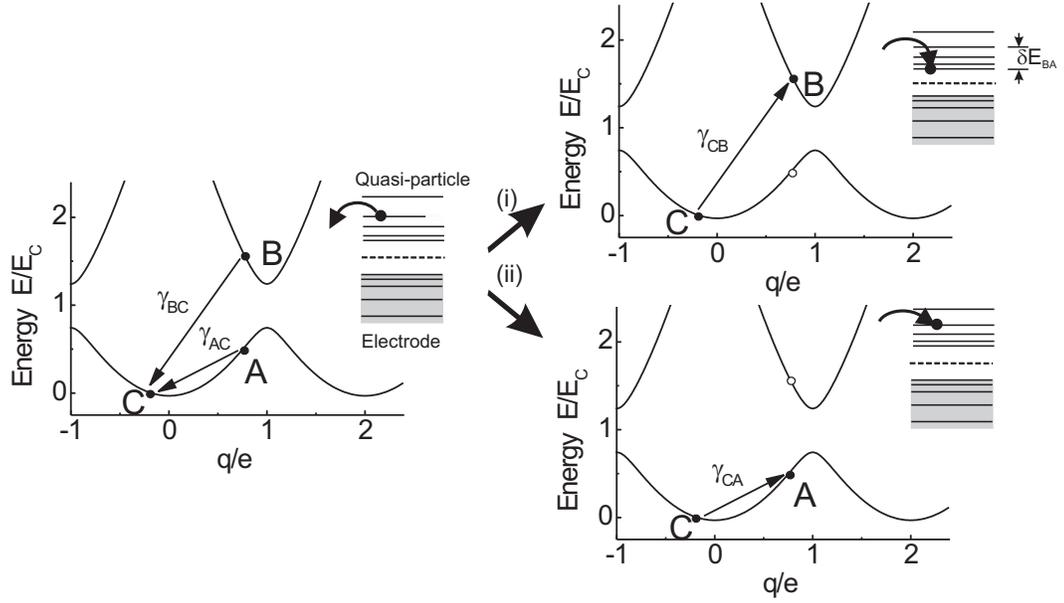}
\caption{\label{fig:fig2} QP-pumping mechanism: In the left panel,
an unpaired non-equilibrium QP tunnels from the outer electrode
onto the island of the qubit in the ground state. This corresponds to the transition $A\rightarrow C$. In the favored
 process (i), the QP tunnels  back with a rate $\gamma_{\rm CB}$ into a
lower energetic state of the outer electrode   and, hence,
transfers
 the energy $\delta E_{BA}$ to the qubit  by exciting it to the upper state. The alternative process (ii), i.e. the tunneling of a QP
back to the outer electrode (with a rate $\gamma_{\rm CA}$  without the qubit
being excited is shown in the panel at the lower right. Due to the somewhat larger density of states for lower energies, the rate $\gamma_{CB}$ is larger than $\gamma_{CA}$.}
\end{figure}
Explaining this mixture effect, we rule out the thermal activation
 of the excited state $B$ at the given base temperature (less than 100 mK) and also the excitation as a result of the Landau-Zener (LZ) tunneling due to the
 rf drive which leads to a periodic passing of the degeneracy point
 (at $\varphi_{\rm dc}=\pi$ and $q=e$) \cite{kong}. This assumption is based on the smooth dependence of the observed effect on the detuning $x=|q-e|/e$, whereas the energy splitting  near the degeneracy point strongly depends on the parameter $x$.
  Instead, we explain the
 observed
mixed state  by a strong coupling of non-equilibrium QPs to the
qubit system, which allows the transfer of energy to the qubit and
thus an excitation of its upper state. This QP-assisted pumping of the
qubit can be considered as a cycle in which an unpaired
non-equilibrium QP tunnels onto the island while the qubit is in
the ground state $A$. As soon as the QP enters the transistor
island, it changes the excess charge and induces in this way an
instantaneous change of the working point (from the even-parity
states $A$ or $B$ to the odd-parity state $C$ with corresponding
rates $\gamma_{\rm AC}$ and of the energy level splitting, see
Fig.\,2). Only when the QP tunnels out to a lower energetic state
of the electrode with a tunneling rate $\gamma_{\rm CB}$, a
transfer of energy to the qubit system with a value $E_{\rm
CB}=E_{\rm C}-E_{\rm B}$ occurs, exciting it to the upper state $B$, whereas a QP
tunneling back to the initial state of the electrode with an
energy difference $\delta E_{\rm CA}=E_{\rm C}-E_{\rm A}$ and corresponding rate
$\gamma_{\rm CA}$ does not induce any excitation of the qubit. The
former process should prevail due to the greater density of the
states that are close to the edge of the QP band in the energy
spectrum. Of course, such non-equilibrium QPs should have a
kinetic energy which is larger at least by the value of $\delta E_{\rm
BA}$ than the value of the superconductor energy gap, in
order to transfer energy to the qubit. This would mean, however,
that a QP having a lower kinetic energy could enter the island as
well, but in that case the QP should leave the island without
exciting the qubit. We assume that QPs having an energy lower than
$\delta E_{\rm BA}$ above the energy gap are available in the
outer electrodes and that their relaxation to the gap edge occurs
both via interaction with the lattice of the electrodes and via
the traveling onto the island and back into the electrodes with
simultaneous excitation of the qubit. Unfortunately, we are unable
to draw conclusions about absolute values of quasiparticle tunneling
rates,
 because in our technique we observe only the averaged values of the phase
 shift $\alpha$.

The QP transitions are described by the tunneling Hamiltonian,
$H_T$, where the Josephson coupling term describing the tunneling
of Cooper pairs is naturally included in the Hamiltonian of the
qubit. Applying the Fermi Golden Rule for calculating the
transition rates of the QP tunneling onto (or off) the island,
we obtain, for example, for the transition $A\rightarrow C$, i.e. when
the qubit is initially in the ground state and a single quasiparticle
tunnels into the island:
\begin{equation}
\gamma_{\rm{AC}}(E_{\rm p}) = 2\pi \sum_{\rm k}|\langle
A,{\rm p}|H_T|C,{\rm k} \rangle|^{2} \delta(E_{\rm p}+\delta E_{\rm AC}-E_{\rm k}),
\label{gamma-rate}
\end{equation}
where the system state is described by the state of the qubit ($A$, $B$ or $C$) and of the QP with the energy $E_{\rm p,k}$, where $\xi_{p,k}$ is the kinetic energy of the QPs with momentum $p,k$ in the lead (or on
the transistor) and $E_{\rm p ,k}=\sqrt{\Delta^2+\xi_{\rm p,k}^2}$, the
corresponding energy. As we have discussed in Ref.~\cite{kong},
the matrix element $|\langle A,{\rm p}|H_T|C,{\rm k} \rangle|^{2}$ is
responsible for the interference effect  occurring  at
phase values $\varphi_{\rm dc}=2\pi n$, $n=0,1,\dots$, when QPs
tunnel onto the island
 both as an
electron-like particle and as a hole-like particle with different
phases for the different trajectories. For these values of the
Josephson phase  a destructive interference takes place having a
suppressing effect on the cyclic qubit excitation. Hence, a
selection rule for the QP transitions between the ground state and
the excited state for certain flux bias values can be deduced
\cite{kong}.

In this paper we address the problem of the energy spectrum of non-equilibrium QPs. Therefore, for our analysis we focus on the results of the meauserements at the special flux bias $\varphi_{\rm dc}=\pi$ (yielding the smallest energy splitting $\delta E_{BA}$), where the selection rule gives a negligible contribution, which cannot be resolved due to the finite value
of the amplitude $a$ of the Josephson phase oscillations, see for more details Ref. \cite{kong}. Therefore, the matrix element is nearly energy-independent for
$\varphi_{\rm dc}=\pi$ and, as a result, one can formulate the
rates for the incoming tunneling events of the quasiparticles,
with $f_{\rm dist}(E_{\rm p})$ corresponding to the filling factor
of the non-equilibrium quasiparticles:
\begin{eqnarray}
\gamma_{\rm AC}& \propto &\int_\Delta^\infty dE_{\rm p}\frac{E_{\rm p}}{\sqrt{E_{\rm p}^2-\Delta^2}}\frac{E_{\rm p}+\delta E_{\rm AC}}{\sqrt{\left(E_{\rm p}+\delta E_{\rm AC}\right)^2-\Delta^2}}f_{\rm dist}(E_{\rm p})\label{r1},\\
\gamma_{\rm BC}& \propto& \int_\Delta^\infty dE_{\rm p}\frac{E_{\rm p}}{\sqrt{E_{\rm p}^2-\Delta^2}}\frac{E_{\rm p}+\delta E_{\rm BC}}{\sqrt{\left(E_{\rm p}+\delta E_{\rm BC}\right)^2-\Delta^2}}f_{\rm dist}(E_{\rm p})\label{r2}.\\
\nonumber\end{eqnarray} This distribution function $f_{\rm dist}$
reflects the occupation number of the QPs for different energy.
The first term in the integrand of formulas (\ref{r1}) and (\ref{r2}) corresponds to the
density of states in the electrodes and the second term to the
density of states in the transistor island. For the tunneling out of
the transistor island, the $\Theta$-function appears because there
are no states to tunnel for a quasiparticle with energy lower than
the threshold energy $E_{\rm k}-\left(\Delta+\delta E_{\rm
CA,CB}\right)$,
\begin{eqnarray}
\gamma_{\rm CA}& \propto &\int_\Delta^\infty dE_{\rm k}\frac{E_{\rm k}}{\sqrt{E_{\rm k}^2-\Delta^2}}\frac{E_{\rm k}+\delta E_{\rm CA}}{\sqrt{\left(E_{\rm k}+\delta E_{\rm CA}\right)^2-\Delta^2}}f_{\rm dist}(E_{\rm k})\Theta\left(E_{\rm k}-\left(\Delta+\delta E_{\rm CA}\right)\right),\\
\gamma_{\rm CB} & \propto &\int_\Delta^\infty dE_{\rm k}\frac{E_{\rm k}}{\sqrt{E_{\rm k}^2-\Delta^2}}\frac{E_{\rm k}+\delta E_{\rm CB}}{\sqrt{\left(E_{\rm k}+\delta E_{\rm CB}\right)^2-\Delta^2}}f_{\rm dist}(E_{\rm k})\Theta\left(E_{\rm k}-\left(\Delta+\delta E_{\rm CB}\right)\right).\\
\nonumber\end{eqnarray}
\begin{figure}
\includegraphics[width=0.89\linewidth]{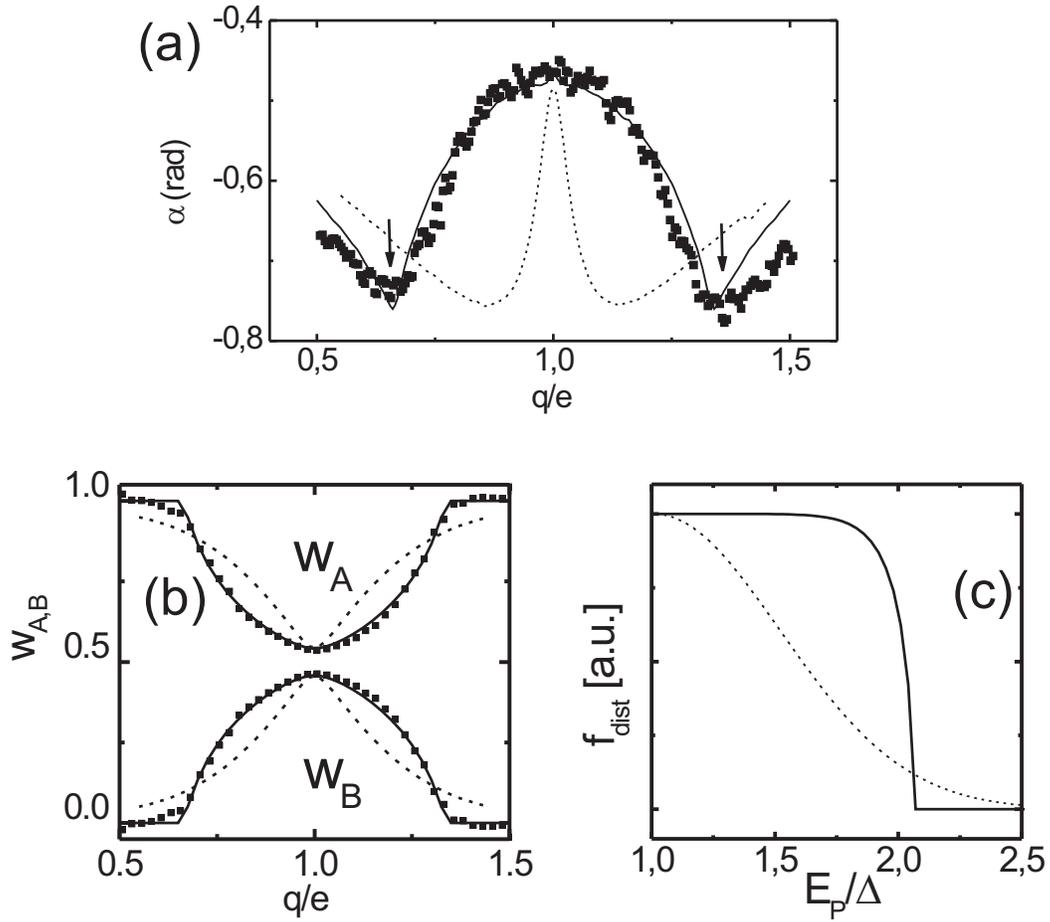}
\caption{\label{fig:fig3} (a) Gate dependence $\alpha(q)$ for a
flux bias $\varphi_{\rm dc}=\pi$; symbols: experimental curve,
solid curve: calculated gate dependence with best-fit spectrum;
dotted curve shows the calculated $\alpha(q)$-curve for a
Gaussian-like spectrum. (b) Occupation probabilities of the ground
state $w_{\rm A}$ and excited state $w_{\rm B}$, yielding the
observed gate dependence. Symbols: reconstructed occupation
numbers from the experimental gate dependence in (a). The solid
curve is calculated using the best-fit spectrum of the
non-equilibrium QPs (solid curve in (c)). The dotted curve is
calculated assuming a Gaussian spectrum of the QPs (dotted curve
in (c)). (c) Energy distribution function of the non-equilibrium
quasiparticles used in the calculation of the QP transition
rates. Solid curve: best-fit spectrum, dotted curve: Gaussian-like
spectrum of the quasiparticles.}
\end{figure}

 We calculate the  steady-state values of the probability weights $w_{A,B}(q)$
 from a system of rate equations
 \begin{eqnarray}
\dot{w}_{\rm A} & = & \gamma_{\rm R}w_{\rm B}+\gamma_{\rm
CA}w_{\rm
C}-\gamma_{\rm AC}w_{\rm A},\\
\dot{w}_{\rm B} & = & -\gamma_{\rm R}w_{\rm B}-\gamma_{\rm
BC}w_{\rm C}+\gamma_{\rm CB}w_{\rm C},\\
\dot{w}_{\rm C} & = &-\gamma_{\rm CB}w_{\rm C}-\gamma_{\rm
CA}w_{\rm C}+\gamma_{\rm BC}w_{\rm B}+\gamma_{\rm AC}w_{\rm A},
\end{eqnarray}
which allows us to estimate the ratio $ \frac{w_{\rm B}}{w_{\rm
A}}\approx \frac{\gamma_{\rm AC}\cdot \gamma_{\rm CB}}{\gamma_{\rm
CA}\cdot \gamma_{\rm BC}}$. Here, we assume that the relaxation
due to quasiparticle tunneling is dominant, i.e.  the rate  of relaxation due to other mechanisms, $\gamma_{\rm R}=\gamma_{\rm BA}$,  
can be neglected.
As we shall see below, the admixture of the excited state is rather strong, which would be impossible if $\gamma_{\rm R}$ played an important role. Such relaxation
also
 includes  the effect of environmental degrees of freedom
 (e.g., flux and gate control lines, external magnetic field, background charge,
 etc.).


 Moreover, the occupation number $w_C$ is negligible, as discussed
 above. The filling factor function  $f_{\rm dist}$ is not
 necessarily normalized, but the corresponding prefactors cancel
 out in the ratio $w_{\rm B}/w_{\rm A}$.

 With the tunneling rates calculated, we are able to model the
peculiar experimental gate modulation (Fig.\,3 (a)) in terms of
occupation numbers of the ground state $w_A$ and the excited state
$w_B$. The experimental values of these quantities are extracted
from Eq.\,(\ref{fa}), fitted subsequently to the measured
$\alpha(q)$-dependence and assume a statistical mixture of the
states $A$ and $B$
\begin{equation}
L^{-1}_{\rm J}(q,\pi)\rightarrow w_A(q)L^{-1}_{\rm
J}(0,q,\pi)+w_B(q)L^{-1}_{\rm J}(1,q,\pi)\label{af}
\end{equation}
(whereby the weight factors are non-negative occupation numbers
which obey the relation $w_A(q)+w_ B(q)=1$).

Hence, we are able to reconstruct the occupation numbers
$w_{A,B}(q)$ of the ground state and of the excited state, see
Fig.\,3 (b). As a result, we find at $q=\pm e$ an increase in the
occupation of the upper state up to $w_A(e)\approx 0.46$, which remains
rather constant in a broad range around this degeneracy point. This
observed steady-state population mirrors the competition between
the rates of quasiparticle tunneling transitions.

As a result of the fitting procedure to the experimental
gate dependence $\alpha(q)$, we can extract some information about the  spectrum of the
non-equilibrium QPs in the electrodes. First, we assumed a
Gaussian-like energy distribution $f_{\rm dist}$ (shown by the dashed line in Fig.\,3 (c)), but such an
approach could not recover the correct shape of neither the
gate curve in Fig.\,3 (a) nor the occupation number dependencies
in Fig.\,3 (b). A reasonable correspondence to the experimental
data was achieved when using a  spectral function $f_{\rm
dist}\propto\sqrt{1-(E_{\rm p}/2.06\Delta)^{20}}$ with a very sharp cut-off for calculating
the ratio $w_{\rm B}/w_{\rm A}$ from the steady-state rate
equations for the tunneling rates. The cut-off energy of  $2.06\Delta_{\rm Al}\approx 430$ $\mu$eV agrees roughly with the turning-points $q=0.58e$, resp. $q=1.42e$ (marked by arrows) of the experimental gate curve in Fig.\,3 (a), from which we can estimate the maximum possible energy transfer $dE$ to the qubit. From the weak-coupling approximation yielding a parabolic shape of the energy bands ($E\approx E_{\rm C}(q/e)^2$ for $E_{\rm J}\ll E_{\rm C}$)  we obtain  $dE=(1.42^2-0.58^2)E_{\rm C}\approx 190$ $\mu$eV, being of the order of $\Delta$. The total energy $E_{\rm P}$ of a quasiparticle with respect to the Fermi level is therefore 
about $2\Delta$. One may speculate about the
origin of the non-equilibrium QPs having such an energy spectrum with an almost equal distribution up to a sharp cut-off energy.
This sharp cut-off energy of approx. $430$
$\mu$eV cannot be explained by the filtering of our signal lines
by ThermoCoax cables located close to the mixing chamber. These cables have a cut-off
frequency of about 1 GHz and the monotonically increased damping at higher frequencies \cite{thermo}. 
Thus, the source of the found QP distribution is still unknown. On the other hand, previous works studying non-equilibrium effects in superconductors and their applications \cite{ilin,howell} have discussed  mechanisms of QP relaxation and pointed out that the relaxation of high-energy QPs is a two-step process. In the first step, the most efficient mechanism of the relaxation of hot QPs is the emission of phonons via inelastic electron-phonon scattering, which enable themselves to break Cooper-pairs due to their short phonon wavelength and create secondary QPs. This happens fast with the electron-scattering time $\tau_{\rm E}$ as the characteristic time scale, e.g $\tau_{\rm E}\propto 10^{-8}$~s \cite{kaplan} for Al at the critical temperature.
This avalanche stops at an average decay energy of $2\Delta$. At this energy, the multiplication of quasiparticles due to the  breaking of Cooper-pairs is  forbidden for  energies lower than $2\Delta$ \cite{kadin} because of energy conservation principle. Instead, QP-QP scattering  leads to a band of QP energies lying between $\Delta$ and $2\Delta$.
In the second step, the decay to thermal equilibrium happens on much longer time scales (the so-called phonon bottleneck). Here, the energy is continually exchanged between the phonon and the QP bath, i.e. the QPs recombine slowly to Cooper pairs under the emission of thermal (long-wave) phonons. Consequently, such a two-step relaxation mechanism might induce the existence of  the deduced energy distribution, as detected in our experiment.

One might utilize this effect for an alternative qubit design which contains electrodes with a lower gap energy than that of aluminium, e.g. titanium, and an Al island. Such a design is similar to the well-known band-gap engineering ($\Delta < \Delta_{\textrm{island}}$) \cite{chargeexp3,tsai} ensuring immediate escape of a QP from the island. In the case of a  sufficiently low energy gap with respect to the energy of the qubit biased in the optimum point $\varphi = 0$ and $q=e$, i.e. $\Delta < E_{J1} + E_{J2}$, the probability of QP-assisted excitation of the qubit is drastically reduced due to the low population of QPs. On the other hand, the process of poisoning  the island without qubit excitation is suppressed due to the selection rule \cite{kong}. In such a regime of the qubit operation the unwanted QP tunneling may be substantially reduced.

In conclusion, we have analyzed the effect of  a statistical
mixing of qubit states related to the energy transfer between
 non-equilibrium quasiparticles and the qubit
 system. In our numerical analysis we have applied a
 rate-equation model to the quasiparticle-induced transitions. Thus, we were able to predict the statistical mixing ratio
 of the qubit states and  to model the experimentally
 observed qubit gate dependence as a result of the qubit pumping. Since our set-up only permitted a continuous readout
 of the, therefore, averaged qubit state, we were unable to provide absolute numbers for the rate of quasiparticle
 tunneling. On the basis of our simulations we can, nevertheless, deduce the energy
 distribution of the non-equilibrium quasiparticles having a
 sharp cut-off with an energy of roughly $2\Delta$.
\begin{acknowledgments}
We would like tothank Michael Wulf for useful discussions. This work was supported by the European Union (project EuroSQIP).
\end{acknowledgments}


\end{document}